\documentclass[twocolumn,showpacs,preprintnumbers,amsmath,amssymb, showkeys]{revtex4}

\usepackage{graphicx}
\usepackage{dcolumn}
\usepackage{bm}
\usepackage{epsf}

\begin{document}


\title{The influence of localized states charging on $1/f^{\alpha}$ tunneling current noise spectrum}

\author{V.\,N.\,Mantsevich}
 \altaffiliation{vmantsev@yahoo.com}
\author{N.\,S.\,Maslova}%
 \email{spm@spmlab.phys.msu.ru}
\affiliation{%
 Moscow State University, Department of  Physics,
119991 Moscow, Russia
}%

\date{\today }

\begin{abstract}
We report the results of theoretical investigations of low
frequency tunneling current noise spectra component
($1/f^{\alpha}$). Localized states of individual impurity atoms
play the key role in low frequency tunneling current noise
formation. It is found that switching "on" and "off" of Coulomb
interaction of conduction electrons with one or two charged
localized states results in power law singularity of low-frequency
tunneling current noise spectrum $1/f^{\alpha}$. Power law
exponent in different low frequency ranges depends on the relative
values of Coulomb interaction of conduction electrons with
different charged impurities.
\end{abstract}

\pacs{71.10.-w, 73.40.Gk, 05.40.-a}
\keywords{D. Non-equilibrium effects; D. Many-particle interaction; D. Tunneling nanostructures}
\maketitle

\section{Introduction}

    In the present work we discuss one of the possible reasons for the
$1/f^{\alpha}$ noise in the STM/STS junctions. We suggest the
theoretical model for tunneling current noise above the flat
surface and above the impurity atoms on semiconductor or metallic
surfaces. We found out that changing of the power law exponent
above the flat surface and above the impurity atom depends on the
parameters of tunneling junction such as tip-sample separation
\cite {Oreshkin}.

Problem of low frequency noise with $1/f^{\alpha}$ spectra
formation in electron devices is one of the most interesting and
important in recent years. Usually the typical approach to
$1/f^{\alpha}$ noise problem consists of "by hand" introducing of
the random relaxation time $\tau_0$ for two-state system with the
probability distribution function $A/\tau_0^\alpha$. Therefore the
averaged over $\tau_0$ noise spectra of two-states system has
power law singularity. But the physical nature and the microscopic
origin of such probability distribution function in general is
unknown.Up to now only the limited number of works was devoted to
the problem of $1/f^{\alpha}$ noise study. The investigations of
the noise in two-level system was carried out in \cite {Galperin}.
Authors studied current noise in a double-barrier
resonant-tunneling structure due to dynamic defects that switch
states because of their interaction with a thermal bath. The time
fluctuations of the resonant level result in low-frequency noise,
the characteristics of which depend on the relative strengths of
the electron escape rate and the defect's switching rate. If the
number of defects is large, the noise is of the $1/f$ type. In
\cite {Levitov} authors studied shot noise in a mesoscopic quantum
resistor. They found correlation functions of all order,
distribution function of the transmitted charge and considered
Pauli principle as the reason for the fluctuations. Altshuler et
al. \cite {Altshuler} studied current fluctuations in a mesoscopic
conductor. They derived a general expression for the fluctuations
in the cylindrical tunneling contact in the presence of a time
dependent voltage. In \cite {Moller} Moller, Esslinger and
Koslowski have investigated the noise of tunneling current at zero
bias voltage. The measurements were carried out at UHV conditions
at base pressure $5\times10^{-11}$ torr. The authors have
demonstrated that at zero bias voltage the $1/f^{\alpha}$
component of noise in the tunneling current vanishes and white
noise becomes dominant. Tiedje {\it et al.} \cite {Tiedje} have
found the $1/f^{\alpha}$ dependence of the current noise in STM
experiments on graphite in ambient conditions. They attributed
this effect to fluctuations induced by adsorbates in tunneling
junction area.

 In \cite{Lozano} the fluctuations of tunneling barrier height have
been investigated. The experiments have been performed under UHV
conditions on graphite and gold samples using PtIr tips. From
these measurements the authors have concluded that the intensity
of barrier height fluctuations correspond to the intensity of
tunneling current $1/f^{\alpha}$ noise in the frequency range from
1 to 100 Hz.

    Our aim is to study one of the possible microscopic origins
of $1/f^{\alpha}$ noise in tunneling contact. We shall analyse
tunneling contact model with localized states and Coulomb
interaction between them which gives us opportunity to describe
the formation of $1/f^{\alpha}$ tunneling current noise spectra.
\section{The suggested model and main results}
 Let's start from the model of two localized states in tunneling
contact. In this case one of the localized states is formed by
impurity atom in semiconductor and the other one by tip apex.

    When electron tunnels in or from localized state, the electron
 filling numbers of localized state rapidly change leading to
 appearance of localized state additional charge and sudden
 switching "on" and "off" Coulomb potential. Electrons in the
 leads feel this Coulomb potential.

The model system (Fig.~1) can be described by hamiltonian
$\Hat{H}$:

$$\Hat{H}=\Hat{H}_{0}+\Hat{H}_{tun}+\Hat{H}_{int}$$
\begin{eqnarray}
&\Hat{H}_{0}&=\sum_{p}(\varepsilon_{p}-eV)c_{p}^{+}c_{p}+\sum_{k}\varepsilon_{k}c_{k}^{+}c_{k}+\sum_{i=1,2}\varepsilon_{i}a_{i}^{+}a_{i}\nonumber\\
&\Hat{H}_{tun}&=\sum_{k,i}T_{ki}c_{k}^{+}a_{i}+\sum_{p,i}T_{pi}c_{p}^{+}a_{i}+T\sum
a_{1}^{+}a_{2}+h.c.\nonumber\\
&\Hat{H}_{int}&=\sum_{k,k^{'}}W_{1}c_{k}^{+}c_{k^{'}}a_{1}a_{1}^{+}+
W_{2}c_{k}^{+}c_{k^{'}}a_{2}a_{2}^{+}\
\end{eqnarray}

 $\Hat{H}_{0}$ describes free electrons in the leads and in
localized states. $\Hat{H}_{tun}$ describes tunneling transitions
between the leads through localized states. $\Hat{H}_{int}$
corresponds to processes of intraband scattering caused by Coulomb
potentials $W_{1}$, $W_{2}$ of localized states charges.

 Operators $c_{k}^{+}(c_{k})$ and $c_{p}^{+}(c_{p})$ correspond to
electrons in the leads and operators $a_{i}^{+}(a_{i})$ correspond
to electrons in the localized states with energy
$\varepsilon_{i}$.

\begin{figure}
\leavevmode\centering{\epsfbox{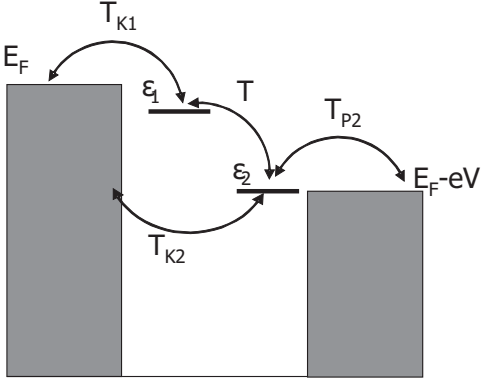}}
 \caption{  Schematic diagram of tunneling processes through states
localized on impurity atom and on the STM tip apex.}
\end{figure}

 Current noise correlation function is determined as:
\begin{eqnarray}
(\hbar/e)^{2}\cdot S(t,t')=<I_{L}(t)\cdot{I_{L}(t')}>=\nonumber\\
=\sum_{k,k^{'},i,j}T_{k}^{2}<c_{k}^{+}(t')a_{i}(t')a_{j}^{+}(t)c_{k_{'}}^{+}(t)>
\
\end{eqnarray}
where
\begin{eqnarray}
I_{L}(t)=\sum_{k}\dot{n}_{k}\cdot
e=(\sum_{k}c_{k}^{+}(t')a_{i}(t')T_{ki}-h.c.)\cdot\frac{e}{\hbar}\
\end{eqnarray}
 The current noise spectra is determined by Fourier
transformation of $S(t,t')$: $S(\omega)=\int S(\tau)d\tau\cdot
e^{i\omega\tau}$.

    We shall use Keldysh diagram technique in our study of low
frequency tunneling current noise spectra \cite {Keldysh}.
Tunneling current noise spectra $S(\omega)$ can be expressed
through Keldysh Green functions.

 For the one localized state in tunneling contact one should put:
$T=0$, $T_{k2}=0$, $\varepsilon_{2}=0$.

 Expression for tunneling current noise spectra without Coulomb
re-normalization of the tunneling vertexes can be found from
diagrams shown on Fig.~2a and has the form:

 \begin{eqnarray}
(&\hbar&/e)^{2}\cdot
S_{0}(\omega)=T_{k1}^{2}\cdot\sum_{k,k_{1}}(G_{kk_{1}}^{<}(\omega)\cdot
 G_{11}^{>}(\omega+\omega')+\nonumber\\
 &+&G_{11}^{<}(\omega)\cdot
G_{kk_{1}}^{>}(\omega+\omega'))
+T_{k1}^{2}\cdot\sum_{k,k_{1}}(G_{k1}^{<}(\omega)\cdot\nonumber\\
&\cdot& G_{k_{1}1}^{>}(\omega+\omega')+G_{k_{1}1}^{<}(\omega)\cdot
G_{k1}^{>}(\omega+\omega'))\
\end{eqnarray}

\begin{figure}
\leavevmode\centering{\epsfbox{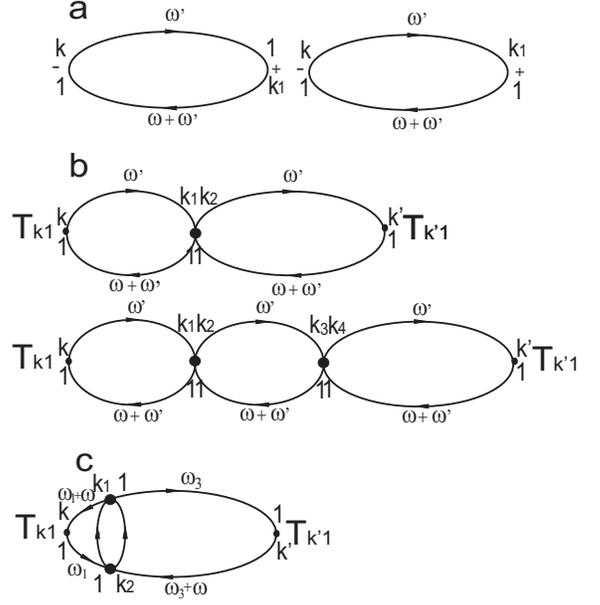}}
 \caption  {Lowest order diagrams contributing to tunneling current noise spectra for one localized state in tunneling contact.
  a) In the absence of Coulomb re-normalization of tunneling vertexes.
  b) In the presence of Coulomb re-normalization of tunneling vertexes - ladder diagrams.
  c) In the presence of Coulomb re-normalization of tunneling vertexes - parquet diagrams.
  Tunneling vertexes are marked by the dot. Coulomb potential is marked by the bold dot. Solid lines correspond to electron Green functions.}
\end{figure}
 Green functions  are evaluated from equations:
\begin{eqnarray}
G_{kk_{1}}^{<}&=&G_{kk_{1}}^{0<}+G_{kk}^{0R}\sum_{k,k^{'}}T_{k^{'}}G_{k^{'}k_{1}}^{<}+G_{kk}^{0<}\sum_{k,k^{'}}T_{k^{'}}G_{k^{'}k_{1}}^{A}\nonumber\\
G_{k1}^{<}&=&G_{kk}^{0R}T_{k1}G_{11}^{<}+G_{kk}^{0<}T_{k1}G_{11}^{A}\\
G_{11}^{<}&=&-2in(\omega)ImG_{11}^{R}(\omega)\nonumber
\end{eqnarray}

 Functions $G_{k_{1}k}^{>}$ and $G_{1k}^{>}$ can be found by
substitution $n_{1}$ on $n_{1}-1$.
\begin{eqnarray}
n_{1}(\omega)=\frac{\gamma_{k1}n_{k}^{o}(\omega)+\gamma_{p1}n_{p}^{o}(\omega)}{\gamma_{k1}+\gamma_{p1}}
\end{eqnarray}
are non-equilibrium localized state filling numbers.

$n_{k}^{o}(\omega)$ and $n_{p}^{o}(\omega)$- equilibrium filling
numbers in the leads. In our model relaxation rates $\gamma_{ki}$,
$\gamma_{pi}$ are determined by electron tunneling transitions
from localized states to the leads $k$ and $p$ continuum states:
\begin{eqnarray}
\sum_{p}T_{pi}^{2}ImG_{pp}^{0R}=\gamma_{pi};
\sum_{k}T_{ki}^{2}ImG_{kk}^{0R}=\gamma_{ki}\
\end{eqnarray}
 For tunneling current noise spectra without
Coulomb re-normalization of tunneling vertexes after substitution
the correspondent Green functions can be written as:
\begin{eqnarray}
&(\label{one} &\hbar/e)^{2}\cdot
S_{0}(\omega)=\gamma_{k1}^{2}\cdot\int
d\omega'ImG_{11}^{R}(\omega')\cdot\nonumber\\
&\cdot& ImG_{11}^{R}(\omega+\omega')\cdot
(n_{1}(\omega+\omega')-1)\cdot\nonumber\\
&\cdot&(n_{1}(\omega')-n_{k}(\omega'))+
n_{1}(\omega')\cdot(n_{1}(\omega+\omega')-\nonumber\\
&-&n_{k}(\omega+\omega')+\gamma_{k1}^{2}\cdot\int
d\omega'ImG_{11}^{R}(\omega')\cdot\nonumber\\
&\cdot&ImG_{11}^{R}(\omega+\omega')\cdot
(n_{k}(\omega+\omega')-1)\cdot n_{1}(\omega')-\nonumber\\
&-&n_{1}(\omega')\cdot(n_{1}(\omega+\omega')-1)-
n_{k}(\omega+\omega')\cdot\
 n_{k}(\omega')+\nonumber\\
&+&n_{k}(\omega')\cdot(n_{1}(\omega+\omega')+ \gamma_{k1}\cdot\int
d\omega'ImG_{11}^{R}(\omega+\omega')\cdot\nonumber\\
&\cdot&(n_{k}(\omega'))\cdot(n_{1}(\omega+\omega')-1)+
ImG_{11}^{R}(\omega')\cdot\nonumber\\
&\cdot&(n_{1}(\omega'))\cdot(n_{k}(\omega+\omega')-1)=\widetilde{S}_{0}\
\end{eqnarray}

 This expression gives us an opportunity to analyze tunneling
current noise spectra for typical values of kinetic parameters of
tunneling contact when localized charge is connected with the tip
apex state $eV=\varepsilon_{1}$. Some low frequency spectra are
shown on Fig.~3a.
\begin{figure*}
\leavevmode\centering{\epsfbox{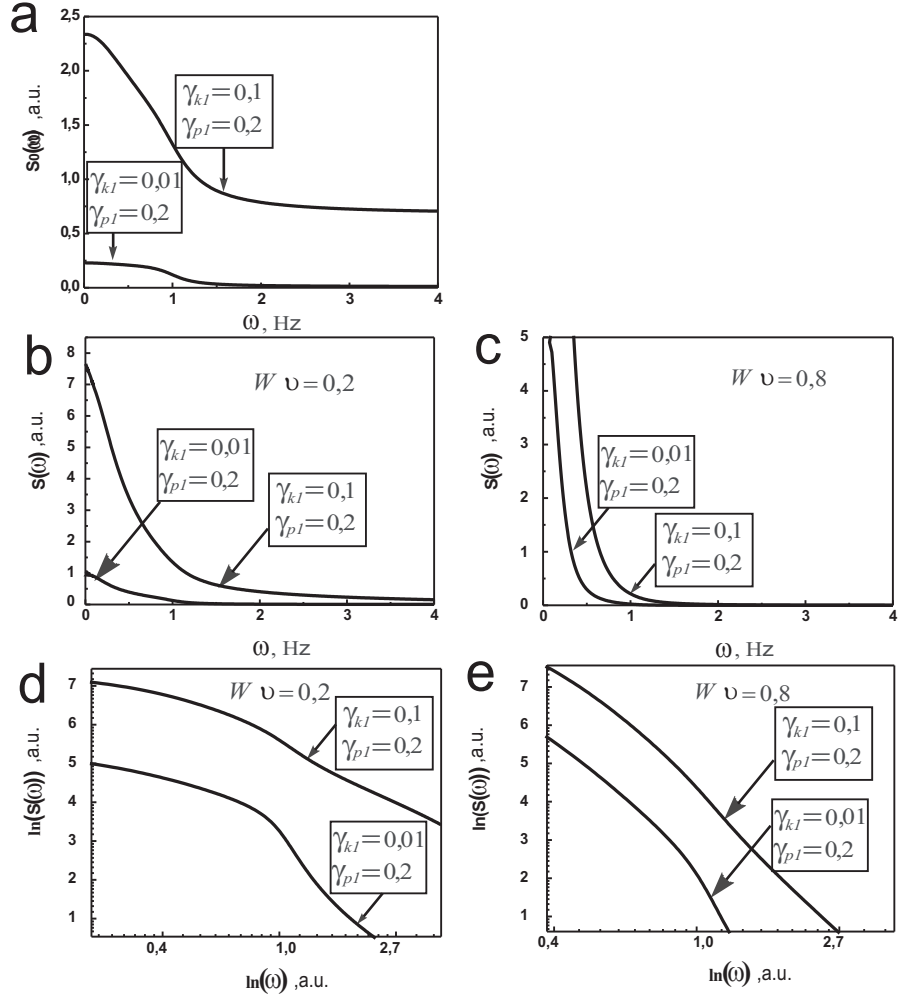}}
 \caption  {  Typical low frequency tunneling current noise spectra for different values of dimensionless kinetic
 parameters for one localized state in tunneling contact ($eV=\varepsilon_1=1$).
  a) In the absence of Coulomb re-normalization of tunneling vertexes.
  b)-c) In the presence of Coulomb re-normalization of tunneling vertexes.
  d)-e) In the presence of Coulomb re-normalization of tunneling vertexes in double logarithmic scale.}
\end{figure*}

    It is clearly evident that when frequency aspire to zero tunneling
current spectra aspire to constant value.

    Now let us consider re-normalization of the tunneling amplitude
and vertex corrections to the tunneling current noise spectra
caused by Coulomb interaction between charged localized state and
the electrons in tunneling contact leads. The result of
re-normalization is shown on Fig.~2b and Fig.~2c. Re-normalization
gives us two types of diagrams contributing to the final tunneling
current noise spectra expression. Ladder diagrams is the most
simple type of diagrams which gives logarithmic corrections to
vertexes. Ladder diagrams give logarithmic divergency at the
threshold voltage $eV=\varepsilon_{1}$ (Fig.~2b). But this is not
the only relevant kind of graphs. We must consider one more type
of graphs (parquet graphs) which gives logarithmically large
contribution to tunneling spectra when $\omega\rightarrow0$ and
$eV=\varepsilon_{1}$ (Fig.~2c). In parquet graphs a new type of
"bubble" appears instead of "dots" in ladder diagrams.
 In this situation one should retain in the n-th order of perturbation
expansion the most divergent terms. For the first time this method
was developed by Dyatlov et. al. \cite {Dyatlov}. It was shown
that for proper treatment of this problem one should write down
integral equations for the so called parquet graphs, which are
constructed by successive substitution the simple Coulomb vertex
for the two types of bubbles in perturbation series. The integral
equations can be solved with logarithmic accuracy, as it was done,
for example, by Nozieres for edge singularities in X-ray
absorption spectra in metals \cite {Nozieres}. The expression for
tunneling current noise spectra after Coulomb re-normalization can
be written as:

\begin{eqnarray}
&(&\hbar/e)^{2}\cdot
S(\omega)=\widetilde{S}_{0}(\omega)+\widetilde{S}_{0}(\omega)\cdot\nonumber\\
&\cdot&((\frac{D^{2}}{(\omega+eV-\varepsilon_{1})^{2}+(\gamma_{k1}+\gamma_{p1})^{2}})^{W\nu}+\nonumber\\
&+&(\frac{D^{2}}{(-\omega+eV-\varepsilon_{1})^{2}+(\gamma_{k1}+\gamma_{p1})^{2}})^{W\nu})\
\end{eqnarray}
where D is the bandwidth for electrons in tunneling contact leads,
, W-Coulomb potential, $\nu$- the equilibrium density of states in
the tunneling contact leads. We consider that localized state is
formed by STM tip apex, so we have to put $eV=\varepsilon_1$. Then
we obtain final expression:

\begin{eqnarray}
(\hbar/e)^{2}\cdot
S(\omega)=\widetilde{S}_{0}(\omega)(1+(\frac{D^{2}}{\omega^{2}+(\gamma_{k1}+\gamma_{p1})^{2}})^{W\nu})\
\end{eqnarray}

 Fig.~3b-3c demonstrate low frequency tunneling current noise spectra
for typical values of dimensionless kinetic parameters. We can see
that re-normalization of tunneling matrix element by switched "on"
and "off" Coulomb potential of charged impurity lead to typical
power law singularity in low frequency tunneling current noise
spectra.

 Some tunneling current noise
spectra in double logarithmic scale for typical values of kinetic
parameters are shown on Fig.~3d-3e. Low frequency tunneling
current noise spectra in double logarithmic scale make it clear
that power law exponent depends only on Coulomb potential of
charged impurity and does not depend on tunneling contact
parameters.

 Now let's describe interaction effects of two localized states in
tunneling contact: surface localized state, formed by impurity
atom, and localized STM tip apex state (Fig.~1). Expression which
describes tunneling current noise spectra without Coulomb
re-normalization can be calculated from graphs, shown in Fig.~4a.
It consists of three parts.

\begin{eqnarray}
\widetilde{S}_{0}(\omega)=\widetilde{S}_{01}(\omega)+\widetilde{S}_{02}(\omega)+\widetilde{S}_{03}(\omega)\
\end{eqnarray}

 $\widetilde{S}_{01}(\omega)$ and $\widetilde{S}_{02}(\omega)$ are rather simple parts
equal to tunneling current noise spectra without Coulomb
re-normalization in the case of one localized state in tunneling
contact Eq.~(\ref{one}). $\widetilde{S}_{03}(\omega)$ is not
trivial, it exists only due to electron tunneling transitions from
one lead to both localized states. Green functions shown on the
graphs are found in \cite {Maslova}. The contribution of
$\widetilde{S}_{01}(\omega)$ is given by graphs with $i=j=1$
(Fig.~4a), $\widetilde{S}_{02}(\omega)$ is described by graph with
$i=j=2$ (Fig.~4a), $\widetilde{S}_{03}(\omega)$ is given by
diagrams with $i\neq j$ (Fig.~4a).
 The final expression for tunneling current noise spectra for two
localized states in tunneling contact without Coulomb
re-normalization of tunneling vertexes after substitution the
correspondent Green functions can be written as:
\begin{figure}
\leavevmode\centering{\epsfbox{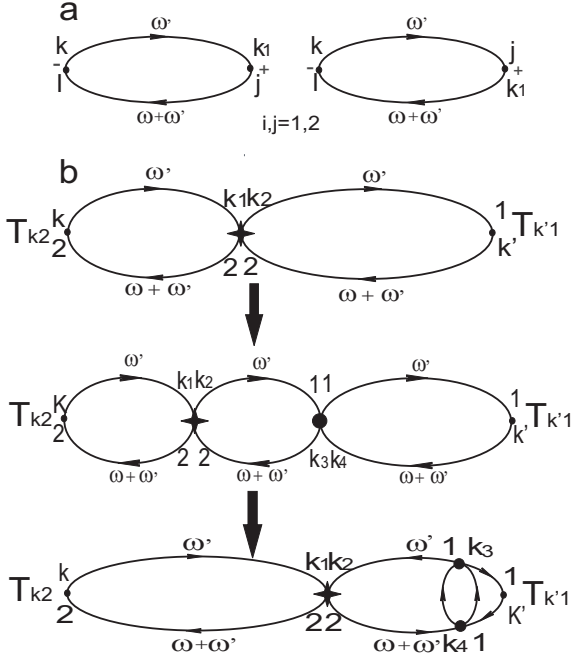}}
 \caption{ Lowest order diagrams contributing to tunneling current noise spectra for two localized states in tunneling contact.
 a) In the absence of Coulomb re-normalization of tunneling vertexes.
 b) In the presence of Coulomb re-normalization of tunneling vertexes.
 Tunneling vertexes are marked by the dot.
 Coulomb potential $W_{1}$ is marked by the bold dot.
 Coulomb potential $W_{2}$  is marked by the star. Solid lines correspond to electron Green functions.}
\end{figure}

\begin{eqnarray}
&(&\hbar/e)^{2}\cdot
S_{03}(\omega)=8\cdot\gamma_{k_{1}}\cdot\gamma_{k_{2}}\cdot \int
d\omega'ImG_{11}^{R}(\omega')\cdot\nonumber\\
&\cdot&ImG_{22}^{R}(\omega+\omega')\cdot
(n_{1}(\omega')\cdot(n_{2}(\omega+\omega')-1)+\nonumber\\&+&
n_{k}(\omega')\cdot(n_{2}(\omega+\omega')-1)+
(n_{1}(\omega')\cdot\nonumber\\
&\cdot&(n_{k}(\omega+\omega')-1)- n_{k}(\omega')\cdot
(n_{k}(\omega+\omega')-1)+\nonumber\\
&+&8\cdot\gamma_{k_{1}}\cdot\gamma_{k_{2}}\cdot\int
d\omega'ImG_{22}^{R}(\omega')\cdot
ImG_{11}^{R}(\omega+\omega')\cdot\nonumber\\
&\cdot&(n_{2}(\omega')\cdot(n_{1}(\omega+\omega')-1)+\nonumber\\
&+&n_{k}(\omega')\cdot(n_{1}(\omega+\omega')-1)+
(n_{2}(\omega')\cdot\nonumber\\
&\cdot&(n_{k}(\omega+\omega')-1)-n_{k}(\omega')\cdot(n_{k}(\omega+\omega')-1)\
\end{eqnarray}

 Non-equilibrium filling numbers $n_{1}$ and $n_{2}$  are determined
from Dyson equations for Keldysh functions $G_{ij}^{<}$, where
$i,j=1,2$ in \cite {Maslova}:

 Green functions $G_{11}^{R}, G_{22}^{R}$ have rather simple form
 and have been also derived in  \cite {Maslova}.
We don't take in account diagrams including Green functions
$G_{12}^{<}, G_{21}^{<} $.  In our case of weak interaction
between localized states $(T<\gamma_{k1}, \gamma_{k2},
\gamma_{p2})$ contribution to the noise spectra of such diagrams
has additional small parameter in comparison with diagrams
depicted on Fig.4a:
\begin{eqnarray}
\frac{T\cdot\gamma_{k1}}{(\varepsilon_{1}-\varepsilon_{2})^{2}+(\gamma_{k1}+\gamma_{k2}+\gamma_{p2})^{2}}\sim
T/\gamma
\end{eqnarray}
 Some typical low frequency tunneling current noise spectra for
different values of dimensionless kinetic parameters without
Coulomb re normalization are shown on Fig.~5a. It is clearly
evident that when frequency aspire to zero tunneling current
spectra aspire to constant value for different dimensionless
kinetic parameters. Now let us consider re-normalization of the
tunneling amplitude and vertex corrections to the tunneling
current spectra caused by Coulomb interaction between both
localized states and tunneling contact leads. Re-normalization
gives us two types of diagrams contributing to the final tunneling
current noise spectra expression similar to types of graphs in the
case of one localized state in tunneling contact. It is necessary
to re-normalize each vertex individually and re-normalize both
vertexes jointly. For each diagram re-normalized by Coulomb
potential of localized state one have to take in to account the
whole series of diagrams including Coulomb interaction with
another localized state (Fig.~4b).

 The final expression for tunneling current noise spectra after
Coulomb re-normalization of tunneling vertexes can be written as:

\begin{eqnarray}
&(&\hbar/e)^{2}\cdot
S(\omega)=\widetilde{S}_{0}(\omega)+\widetilde{S}_{01}(\omega)\cdot\nonumber\\
&\cdot&((\frac{D^{2}}{(\omega+E_{1})^{2}+\Gamma_{2}^{2}})^{W_{1}\nu}+
(\frac{D^{2}}{(\omega+E_{2})^{2}+\Gamma_{1}^{2}})^{W_{1}\nu})+\nonumber\\
&+&\widetilde{S}_{02}(\omega)\cdot
((\frac{D^{2}}{(-\omega+E_{1})^{2}+\Gamma_{1}^{2}})^{W_{2}\nu}+\nonumber\\
&+&(\frac{D^{2}}{(-\omega+E_{2})^{2}+\Gamma_{2}^{2}})^{W_{2}\nu})+
\widetilde{S}_{03}(\omega)\cdot\\
&\cdot&((\frac{D^{2}}{(\omega+E_{1})^{2}+\Gamma_{2}^{2}})^{W_{1}\nu}+
(\frac{D^{2}}{(\omega+E_{2})^{2}+\Gamma_{1}^{2}})^{W_{1}\nu})\cdot\nonumber\\
&\cdot&((\frac{D^{2}}{(-\omega+E_{1})^{2}+\Gamma_{1}^{2}})^{W_{2}\nu}+
(\frac{D^{2}}{(-\omega+E_{2})^{2}+\Gamma_{2}^{2}})^{W_{2}\nu})\nonumber
\end{eqnarray}

where,

\begin{eqnarray}
E_{1}&=&eV-\frac{\varepsilon_{1}+\varepsilon_{2}}{2}+\frac{\sqrt{(\varepsilon_{1}-\varepsilon_{2})^{2}+4T^{2}}}{2}\nonumber\\
E_{2}&=&eV-\frac{\varepsilon_{1}+\varepsilon_{2}}{2}-\frac{\sqrt{(\varepsilon_{1}-\varepsilon_{2})^{2}+4T^{2}}}{2}\\
\Gamma_{1}&=&\frac{(\varepsilon_{1}-\varepsilon_{2})(\gamma_{k2}+\gamma_{p2}-\gamma_{k1})}{2\sqrt{(\varepsilon_{1}-\varepsilon_{2})^{2}+4T^{2}}}+\frac{\gamma_{k2}+\gamma_{p2}-\gamma_{k1}}{2}\nonumber\\
\Gamma_{2}&=&\frac{(\varepsilon_{1}-\varepsilon_{2})(\gamma_{k2}+\gamma_{p2}-\gamma_{k1})}{2\sqrt{(\varepsilon_{1}-\varepsilon_{2})^{2}+4T^{2}}}-\frac{\gamma_{k2}+\gamma_{p2}-\gamma_{k1}}{2}\nonumber
\end{eqnarray}

 When the applied bias voltage become close to $\varepsilon_{2}$ we obtain low frequency tunneling spectra. So we
have to put $eV=\varepsilon_{1}=\varepsilon_{2}$. We obtain the
expression:

\begin{eqnarray}
&(&\hbar/e)^{2}\cdot
S(\omega)=\widetilde{S}_{0}(\omega)+\widetilde{S}_{01}(\omega)\cdot(
(\frac{D^{2}}{\omega^{2}+\Gamma_{1}^{2}})^{W_{1}\nu}+\nonumber\\
&+&(\frac{D^{2}}{\omega^{2}+\Gamma_{1}^{2}})^{W_{1}\nu})+
\widetilde{S}_{02}(\omega)\cdot((\frac{D^{2}}{\omega^{2}+\Gamma_{1}^{2}})^{W_{2}\nu}+\nonumber\\
&+&(\frac{D^{2}}{\omega^{2}+\Gamma_{1}^{2}})^{W_{2}\nu})+
\widetilde{S}_{03}(\omega)\cdot((\frac{D^{2}}{\omega^{2}+\Gamma_{1}^{2}})^{W_{1}\nu}+\\
&+&(\frac{D^{2}}{\omega^{2}+\Gamma_{1}^{2}})^{W_{1}\nu})\cdot
((\frac{D^{2}}{\omega^{2}+\Gamma_{1}^{2}})^{W_{2}\nu}+(\frac{D^{2}}{\omega^{2}+\Gamma_{1}^{2}})^{W_{2}\nu})\nonumber
\end{eqnarray}
\begin{figure}
\leavevmode\centering{\epsfbox{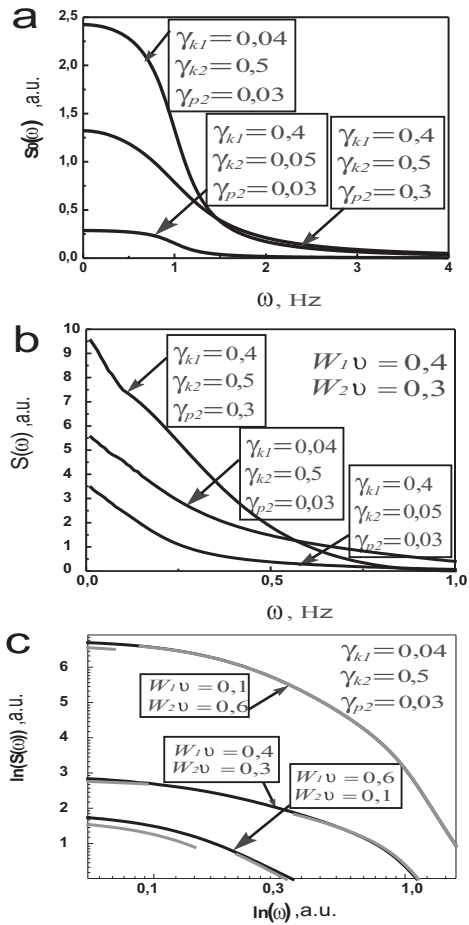}}
 \caption{  Typical low frequency tunneling current noise spectra for different values of dimensionless kinetic
 parameters for two localized states in tunneling contact ($eV=\varepsilon_1=\varepsilon_2=1$).
  a) In the absence of Coulomb re-normalization of tunneling vertexes.
  b) In the presence of Coulomb re-normalization of tunneling vertexes.
  c) In the presence of Coulomb re-normalization of tunneling vertexes in double logarithmic scale. Black lines correspond to tunneling current noise spectra in
  double logarithmic scale. Gray lines correspond to the terms which produce the most strong logarithmic singularity and approximate
  full spectra in the best way.}
\end{figure}

    First of all we consider the situation when both localized
states acquire positive charge.
 Fig.~5b demonstrate low frequency tunneling current noise spectra
for typical values of dimensionless kinetic parameters. We can see
that re-normalization of tunneling matrix element by switched "on"
and "off" Coulomb potential of charged impurities lead to typical
power law singularity in low frequency tunneling current noise
spectra. Tunneling current noise spectra in double logarithmic
scale demonstrate frequency regions where every part of final
expression which include different power law exponent, approximate
noise spectra in the best way (Fig.~5c).
 Let's analyse tunneling current spectra
shown on Fig.~5c. Contribution to tunneling current spectra at low
(zero) frequency is always determined by the term which produces
the most strong of logarithmic singularity, determined by the sum
of localized states Coulomb potentials at any values of
dimensionless tunneling rates of tunneling contact.
 In the case of $W_{1}>>W_{2}$
(Coulomb potential of charged impurity atom is larger than Coulomb
potential of charged tip apex localized state) with frequency
increasing the main contribution to tunneling current noise
spectra is given by the term depending on Coulomb potential of
localized state caused by impurity atom $(W_{1})$ at any
dimensionless kinetic parameters of tunneling contact.
 If  Coulomb potential of
charged tip apex localized state is larger than Coulomb potential
of charged impurity atom $(W_{2}>>W_{1})$ with frequency
increasing the main contribution to tunneling current noise
spectra is given by the term depending on Coulomb potential of tip
apex localized state $(W_{2})$ at any dimensionless kinetic
parameters of tunneling contact.
 When  Coulomb potential of charged tip apex localized
state is similar to Coulomb potential of charged impurity atom
$(W_{1}\sim W_{2})$ with frequency increasing the main
contribution to tunneling current noise spectra can be given both
by the term depending on Coulomb potential of tip apex localized
state $(W_{2})$ and by the item depending on Coulomb potential of
localized state caused by impurity atom $(W_{1})$ at any
dimensionless kinetic parameters of tunneling contact. In this
case the replacement of dominating term can take place. Simple
estimation gives an opportunity to determine the validity of
obtained expressions. For typical typical parameters of tunneling
junction model in the zero frequency region power spectrum of
tunneling current corresponds to experimental results. Power
spectrum on zero frequency has the form:
\begin{eqnarray}
S(0)\approx(\gamma_{eff}\cdot
e/\hbar)^{2}\cdot(D/\gamma_{eff1})^{\nu\cdot W}\cdot
(1/\bigtriangleup\omega)\
\end{eqnarray}
For typical $\gamma_{eff}$, $\gamma_{eff1}$ $\approx10^{-13}$,
$D\approx10$, $W\approx0,5$ $S(0)\approx10^{-18} A^{2}/Hz$ \cite
{Oreshkin}.

\section{Conclusion}
  The microscopic theoretical approach describing tunneling current noise spectra taking in
account many-particle interaction was proposed. When electron
tunnels to or from localized state the charge of localized state
rapidly changes. This results in sudden switching on and off of
additional Coulomb potential in tunneling junction area, and leads
to typical power law dependence for low frequency tunneling
current noise spectra. In the case of two interacting positively
charged localized states the power law exponent at low frequency
is determined by the sum of Coulomb potentials. It was shown that
if one of the Coulomb potentials strongly exceeds the other one
this potential determine tunneling current noise spectra at low
frequency with increasing of frequency.
 If $W_{1}\sim W_{2}$, tunneling current noise
spectra can be determined both by Coulomb potential $W_{1}$, and
by Coulomb potential of charged tip apex state $W_{2}$, depending
on dimensionless kinetic parameters of tunneling contact.
    When localized states acquire charges of opposite signs
tunneling current noise spectra in the low frequency region are
approximated by the term depending on maximum positive value of
Coulomb potentials $W_{1}$, $W_{2}$, $W_{1}-W_{2}$ or
$W_{2}-W_{1}$.

 We are grateful to A.I. Oreshkin and S.V. Savinov for discussions
 and useful remarks.

This work was partially supported by RFBR grants ¹ 06-02-17076-a,
¹ 06-02-17179-a, ¹ 05-02-19806-MF and the Council of the President
of the Russian Federation for Support of Young Scientists and
Leading Scientific School  ¹ NSh-4599.2006.2 and ¹
NSh-4464.2006.2.


\pagebreak

\end{document}